\documentstyle[12pt,epsf]{article}

\newcommand{\wt}{\widetilde}

\newcommand{\ol}{\overline}

\def\diag{\mathop{\rm diag}\nolimits}

\def\str{\mathop{\rm str}\nolimits}

\begin{document}

\baselineskip=18pt plus 0.2pt minus 0.1pt

\begin{titlepage}
\title{
\hfill\parbox{4cm}
{\normalsize YITP-98-43\\{\tt hep-th/9807179}\\July 1998}\\
\vspace{2.5cm}
Supersymmetries and BPS Configurations\\
on Anti-de Sitter Space
}
\author{
Yosuke Imamura\thanks{{\tt imamura@yukawa.kyoto-u.ac.jp}}
%{}\thanks{
%Supported in part by Grant-in-Aid for Scientific
%Research from Ministry of Education, Science and Culture
%(\#9110).}
\\[7pt]
{\it Yukawa Institute for Theoretical Physics,}\\
{\it Kyoto University, Sakyo-ku, Kyoto 606-8502, Japan}
}
\date{}

\maketitle
\thispagestyle{empty}

\begin{abstract}
\normalsize
We study supersymmetry breaking due to the presence of branes
on anti-de Sitter space
and obtain conditions for brane orientations not to break
too many supersymmetries.
Using the conditions,
we construct a brane configuration corresponding to a baryon
in large $N$ gauge theory,
and it is shown that
the baryon is a marginal bound state of quarks
as is expected from supersymmetry.
\end{abstract}

\vspace{3cm}


\end{titlepage}

%%%%%%%%%%%%%%%%%%%%%%%%%%%%%%%%%%%%%%%%%%%%%%%%%%%%%%%%%%%%%%%%
\section{Introduction}
Recently, a baryon configuration
in the context of AdS/CFT correspondence
\cite{Maldacena,WilsonLoop,Holography,Klebanov,GlueBall,nunes,thermal}
has been suggested\cite{WittenBaryon,GrossOoguri}.
Its properties were studied in \cite{Yan,Ima}
and it has been shown that the baryon mass is quarter of
the sum of constituent quark mass.
However, this is very strange because supersymmetries guarantee
that BPS-states cannot fall into lower energy states,
and truly (non-marginal) bound states are forbidden.
The purpose of this paper is to resolve this contradiction
by showing that the configuration is marginal
contrary to the previous result.

The baryon configuration consists of D5-brane wrapped around
$S^5$ and $N$ fundamental strings with the same orientation
attached on the D5-brane.
The wrapped D5-brane feels two forces in the opposite direction to each other,
namely, gravitational force
from $N$ overlapped D3-branes and tension of $N$ fundamental strings
attached on it.
In \cite{Yan,Ima},
it has been shown that the string tension is four times of the gravitational force
and this unbalance is the origin of the nonzero binding energy of the baryon.
However,
there is one doubtful point in the treatment in \cite{Yan,Ima}.
They neglect a deformation of the D5-brane due to the string tension
and energy of an electric field on the D5-brane.
As is mentioned in \cite{Ima},
if $N$ strings are distributed on $S^5$ uniformly,
the treatment in \cite{Yan,Ima} is justified
because string tension acts on the D5-brane almost uniformly and the D5-brane
is not deformed,
and the charges of string end points are almost locally canceled by
Chern-Simons coupling to R-R four form field.
Furthermore, because this configuration breaks all supersymmetries,
there is no inconsistency even if truly bound states appear.
On the other hand, in a supersymmetric configuration,
in which all of $N$ strings lie on one point on $S^5$,
the charge of the string end points
and the string tension pulling the D5-brane
concentrate at one point on the D5-brane.
So the brane may be deformed from $S^5$ and the electric field
on the brane may not be negligible.
Therefore, to obtain the correct energy of the configuration,
we should use a framework given in \cite{BIstring1,BIstring2},
which starts from the Born-Infeld action of D-branes.

In the present paper, we construct the baryon configuration
as a classical solution for Born-Infeld action.
To obtain the solution,
we work out conditions for a brane orientation
not to break too many global supersymmetries
on $AdS_5\times S^5$.
We refer the condition by `BPS condition'.
Using them, we can easily obtain the correct value of the energy of
the baryon configuration.
We find that it is equal to the total energy of $N$ quarks
and it is confirmed that the baryon state is a marginal bound state
as is expected from supersymmetry.

It is instructive to observe properties of a configuration
in Type-I' theory, which is similar to the baryon configuration.
Type-I' theory contains $16$ D8-branes and two orientifold 8-planes.
Let us consider a configuration
in which seven of 16 D8-branes are on the top of the left one
of the two orientifold 8-planes,
eight D8-branes are overlapped on the right 8-plane,
and the last D8-brane lies between them.
The right side of the last D8-brane is a massless region
where spacetime is flat and dilaton vev is constant,
while the left side is called a `massive' region with curved background
\cite{78brane,massiveIIARomans}.
If we put a D-particle in the massless region,
it is stable and is a BPS configuration.
On the other hand, a D-particle in the massive region is
unstable because the background space-time is curved
and the D-particle feels the gravitational force.
To cancel it, we should introduce a fundamental string
stretched between the D-particle and the D8-brane at the boundary
of the two regions.
The string is also regarded as
what is created by Hanany-Witten effect
when the D-particle pass through the D8-brane at the boundary
from the massless region to the massive region.
The obtained configuration is stable and is a BPS state.

Furthermore,
if we compactify the configuration on $T^n$ along the eight branes
and carry out T-duality transformation along all directions of the $T^n$,
we get Type-II theory
(depending on $n$ is even or odd,
we get Type-IIA theory or Type-IIB theory.)
compactified on $T^{n+1}/{\bf Z}_2$,
where ${\bf Z}_2$ act on all cycles of $T^{n+1}$.
Via the T-duality,
the D8-branes and the D-particle are transformed into D$(8-n)$-branes and
a D$n$-brane, respectively.
The D$n$-brane is wrapped around a $T^n\subset T^{n+1}/{\bf Z}_2$
and is connected with one of the D$(8-n)$-branes by a string.
The vev of zero form field strength in the original theory,
which is regarded as a quantized cosmological
constant in the massive region,
is transformed into $n$-form field strength of R-R $n-1$ form potential
which is magnetically coupled to the D$(8-n)$-branes.
In this T-dualized picture, the D$n$-brane feels gravitational force
and string tension,
and this configuration is quite similar to the baryon configuration
in $AdS_5\times S^5$.
%%%%%%%
If the string coupling $g_{\rm str}$ is large,
a deformation of the D$(8-n)$-brane due to
the string tension and
an energy of electric field on the D$(8-n)$-brane are also large.
However, even if we neglect them, we get correct energy of the configuration
because a `BPS condition' guarantees the
balance between the energy of the electric field and
decreasing of area of the D-brane due to the deformation.
Therefore, we may neglect the deformation and the gauge field.

By analogy with this,
one might think that
the ignorance would be justified
by means of supersymmetry also in other situations.
However, as we will show later,
this conjecture does not correct for $AdS$ background.

%%%%%%%%%%%%%%%%%%%%%%%%%%%%%%%%%%%%%%%%%%%%%%%%%%%%%%
\section{Branes and Supersymmetries in Type-IIB Theory}

Type-IIB string theory has $32$ supercharges which are combined into
two left handed Majorana-Weyl spinors $\eta_L$ and $\eta_R$.
If we adopt a signature of space-time
$\eta_{MN}=\diag(-,+,\ldots,+)$,
the charge conjugation matrix $C$ satisfies the following relations:
\begin{equation}
C\Gamma_M=-\Gamma_M^TC,\quad 
C^T=-C.
\end{equation}
The charge conjugation $\psi_c$ of a spinor $\psi$ is given by
\begin{equation}
\psi^TC=\ol\psi,
\end{equation}
where Dirac conjugate is defined by $\ol\psi=\psi^\dagger\Gamma^0$.
Two supersymmetry parameters $\eta_L$ and $\eta_R$ satisfy both
the Majorana condition:
\begin{equation}
\eta_L^TC=\ol\eta_L,\quad
\eta_R^TC=\ol\eta_R,
\end{equation}
and the Weyl condition:
\begin{equation}
\Gamma^{11}\eta_L=+\eta_L,\quad
\Gamma^{11}\eta_R=+\eta_R.
\label{Weylcond}
\end{equation}
(The subscripts $L$ and $R$ indicate the supersymmetries
come from left and right movers of closed strings,
and $\eta_L$ and $\eta_R$ have the same space-time chirality.)
Because left mover and right mover are connected by Dirichlet or
Neumann boundary conditions
at the end points of open strings attached on D-branes,
half of supersymmetries are broken on D-branes.
In the case of vanishing $U(1)$ gauge field on D-branes,
the parameters of unbroken supersymmetries should satisfy
the following
relations between $\eta_L$ and $\eta_R$:
\begin{equation}
\mbox{%
\begin{tabular}{crr}
D-string  & $\Gamma^{01}\eta_L=+\eta_R$, &
            $\Gamma^{01}\eta_R=+\eta_L$, \\[1ex]
D3-brane  & $\Gamma^{0123}\eta_L=+\eta_R$, &
            $\Gamma^{0123}\eta_R=-\eta_L$, \\[1ex]
D5-brane  & $\Gamma^{012345}\eta_L=+\eta_R$, &
            $\Gamma^{012345}\eta_R=+\eta_L$,
\end{tabular}}
\label{Rcharged}
\end{equation}
while NS-NS charged objects do not connect $\eta_L$ and $\eta_R$:
\begin{equation}
\mbox{%
\begin{tabular}{crr}
F-string  & $\Gamma^{01}\eta_L=+\eta_L$, &
            $\Gamma^{01}\eta_R=-\eta_R$, \\[1ex]
NS5-brane & $\Gamma^{012345}\eta_L=+\eta_L$, &
            $\Gamma^{012345}\eta_R=-\eta_R$.
\end{tabular}}
\label{NScharged}
\end{equation}
(In eqs.(\ref{Rcharged}) and (\ref{NScharged}),
we suppose that each $p$-brane lies along $x^0,\ldots,x^p$.)
For the following arguments, it is convenient to
combine two Majorana-Weyl spinors into complex Weyl spinors:
\begin{equation}
\eta^\pm=\eta_L\pm i\eta_R.
\end{equation}
Then, eqs.(\ref{Rcharged}) and (\ref{NScharged})
are recombined into the following constraints:
\begin{equation}
\begin{array}{ccccc}
\mbox{D-string} & \mbox{D3-brane} & \mbox{D5-brane}
 & \mbox{F-string} & \mbox{NS5-brane} \\[1ex]
\Gamma^{01}\eta^\pm=\pm i\eta^\mp, &
\Gamma^{0123}\eta^\pm=\mp i\eta^\pm, &
\Gamma^{012345}\eta^\pm=\pm i\eta^\mp, &
\Gamma^{01}\eta^\pm=\eta^\mp, &
\Gamma^{012345}\eta^\pm=\eta^\mp.
\end{array}
\label{flatbps}
\end{equation}
By comparing the constraint for D-strings and the one for fundamental strings
in (\ref{flatbps}),
we can read the fact that S-duality transformation
is achieved by a phase rotation of the parameters $\eta^\pm$:
\begin{equation}
\eta^\pm\rightarrow\exp\left(\pm\frac{\pi i}{4}\right)\eta^\pm.
\end{equation}
In the expression (\ref{flatbps}), we supposed that the theta angle vanishes.
For a generic value of the theta angle and an arbitrary string charge $(p,q)$,
we can obtain the constraint
from the one for fundamental strings by a phase rotation \cite{ortin}:
\begin{equation}
\eta^\pm\rightarrow\exp\left(\pm\frac{i}{2}\arg(p+q\tau)\right)\eta^\pm,
\label{U1S}
\end{equation}
where $\tau=\theta/2\pi+i/g_{\rm str}$.
Identically, we can obtain the constraint for $(p,q)$ five-branes
from the one for D5-branes.

%%%%%%%%%%%%%%%%%%%%%%%%%%%%%%%%%%%%%%%%%%%%%%%%%%%%%%
\section{Anti-de Sitter Space}
The metric of an extremal three brane lying along $y_{0,\ldots,3}$ is
\begin{equation}
ds^2=H^{-1/2}({\bf y})\sum_{\mu=0}^3dy_\mu^2+H^{1/2}({\bf y})d{\bf y}^2,\quad
H({\bf y})=1+\frac{r_0^4}{r^4},
\label{D3metric}
\end{equation}
where ${\bf y}=(y_4,y_5,\ldots,y_9)$ and $r=|{\bf y}|$.
The radius $r_0$ of the horizon is connected with
a R-R charge $N\in{\bf Z}$ of the three brane by a relation:
\begin{equation}
r_0^4=\frac{2\kappa^2}{4c_5}NT_{\rm D3}=4\pi l_s^4g_{\rm str}N,
\label{r0is}
\end{equation}
where $T_{D3}=1/(2\pi)^3l_s^4g_{\rm str}$ is a tension
of extremal D3-brane with unit charge,
$c_5=\pi^3$ is a volume of the five dimensional unit sphere
and $2\kappa^2=(2\pi)^7l_s^8g_{\rm str}^2$ is Newton's constant.
In this paper, we adopt a convention
in which the string tension is $T_{FS}=1/(2\pi l_s^2)$ and
the string coupling constant $g_{\rm str}$ is
transformed into $1/g_{\rm str}$ by S-duality.
The three brane solution consists of a flat region ($r_0\ll r$) and a
throat near the horizon ($0\leq r\ll r_0$).
In the near horizon region, eq.(\ref{D3metric})
is reduced to $AdS_5\times S^5$ metric:
\begin{equation}
ds^2=\frac{r^2}{r_0^2}\sum_{\mu={\ol1}}^{\ol4}dy_\mu^2
    +\frac{r_0^2}{r^2}dr^2+r_0^2d\Omega_5^2.
\label{near}
\end{equation}
(From now on, we label the coordinates on $AdS_5$ by overlined numbers
$\ol1,\ol2,\ldots$, in principle, while the coordinates on
$S^5$ are labeled by usual numbers $1,2,\ldots$.)
This manifold has an isometry $SO(2,4)\times SO(6)$.
To make it clear, it is convenient to
represent $AdS_5$ as a (pseudo-)sphere
in a flat six-dimensional space with signature $(-,+,+,+,+,-)$:
\begin{equation}
-x_{\ol1}^2+x_{\ol2}^2+x_{\ol3}^2+x_{\ol4}^2+x_{\ol5}^2-x_{\ol6}^2=-r_0^2.
\label{pseudosphere}
\end{equation}
To make contact with the expression (\ref{near}),
we should introduce a coordinate $r$ as a light cone coordinate
and we should rescale $x_{\ol1},\ldots,x_{\ol4}$ by $r_0/r$:
\begin{equation}
r=x_{\ol5}+x_{\ol6},\quad
y_{\ol a}=\frac{r_0}{r}x_{\ol a},
\quad
(\ol a=\ol1,\ol2,\ol3,\ol4),
\end{equation}
and then, the $AdS_5$ part of the metric (\ref{near}) is reproduced.
From the viewpoint of a four dimensional field theory
on the $AdS_5$ boundary at the opening of the throat
\begin{equation}
x_{\ol5}+x_{\ol6}\sim r_0,
\label{openning}
\end{equation}
the isometry $SO(2,4)$ of $AdS_5$ is
regarded as a conformal group.
It consists of Lorentz rotations $M_{\ol a\ol b}$,
parallel transports $P_{\ol a}$, conformal boosts $K_{\ol a}$ and
a dilatation $D$, where indices $\ol a$ and $\ol b$ run over
 $\ol1$, $\ol2$, $\ol3$ and $\ol4$.
Poincar\'e group consisting of $M_{\ol a\ol b}$ and $P_{\ol a}$ is identified
with a subgroup of $SO(2,4)$ which does not
move the boundary plane (\ref{openning}).
Using this fact,
we can establish a correspondence between generators $T_{\ol a\ol b}$ of
$SO(2,4)$ and ones of conformal group as follows:
\begin{equation}
T_{\ol a\ol b}\sim M_{\ol a\ol b},\quad
T_{+\ol a}\sim P_{\ol a},\quad
T_{-\ol a}\sim K_{\ol a},\quad
T_{+-}\sim D,
\end{equation}
where subscripts $\pm$ mean light cone coordinates $x_{\pm}=x_{\ol5}\pm x_{\ol6}$.
By adding rigid supersymmetries $Q$, superconformal symmetries $S$
and $SO(6)$ R-symmetry transformation,
we can extend the conformal group $SO(2,4)$ to a superconformal group
$SU(2,2|4)$.
Concerning the fermionic symmetries,
we discuss them in section \ref{global}.

%%%%%%%%%%%%%%%%%%%%%%%%%%%%%%%%%%%%%%%%%%%%%%%%
\section{Brief Review about Harmonics on $S^d$}

The transformation law of gravitini $\psi^\pm_M$ in type-IIB theory
under local supersymmetry transformation is\cite{IIBequation}
\begin{equation}
\delta\psi^\pm_M=D_M\eta^\pm
             \mp\frac{ia}{5!}
             F_{M_1\ldots M_5}
             \Gamma^{M_1\ldots M_5}
             \Gamma_M\eta^\pm
             +\cdots,
\label{delpsi}
\end{equation}
where $a$ is a constant.
In (\ref{delpsi}), we omit some terms which vanish in $AdS_5\times S^5$ background.
To get global supersymmetries on $AdS_5\times S^5$,
it is necessary to solve a differential equation $\delta\psi_M^\pm=0$.
This equation is easily solved by using spherical harmonics on $\cal M$
where $\cal M$ is one of $AdS_5$ and $S^5$.
In this section, we give a brief review of spherical harmonics
as a preparation for the following sections.

In this section, we suppose ${\cal M}=S^d$.
A generalization to the case of ${\cal M}=AdS_d$ is straightforward.
A $d$ dimensional sphere $S^d$ is represented as a coset space $H\backslash G$
of an isometry group $G=SO(d+1)$ over a local rotation group $H=SO(d)$.
Let $T_{IJ}$ ($I,J=1,\ldots d,d+1$) denote generators of $G$
normalized by
\begin{equation}
\exp(2\pi T_{IJ})=1.
\label{Tnormal}
\end{equation}
Let us assume $H$ is a subgroup of $G$ generated
by $T_{ab}$ ($a,b=1,\ldots d$).
A point $x$ on the manifold ${\cal M}=H\backslash G$
is identified with a set $Hg$
where $g$ is an element of $G$.
The isometry of $\cal M$ is realized by an action of $G$ from the right side
$x\rightarrow x'=xg$.

To express tensor and spinor fields,
we should introduce a vielbein on $\cal M$.
It is achieved by defining a coordinate $\xi^a$ on a tangent space
at a point $x_0\in H\backslash G$ by
\begin{equation}
x=H(1+\xi^aT_{d+1,a})\Sigma(x_0),
\label{localcoord}
\end{equation}
where $x$ is a point near $x_0$
and $\Sigma$ is a mapping from $\cal M$ into $G$
satisfying $H\Sigma(x_0)=x_0$,
namely, a section of the fiber bundle $G$ over $\cal M$.
The definition (\ref{localcoord}) of the local coordinate $\xi^a$
has an ambiguity concerning
the choice of the section $\Sigma$.
It is regarded as an ambiguity concerning the choice of a vielbein.
In fact, if we pick two sections $\Sigma(x)$ and $\Sigma'(x)$,
there is a mapping $h(x)$ from $\cal M$ into $H$
which relates two sections by $\Sigma'(x)=h(x)\Sigma(x)$.
Inserting $\Sigma(x_0)=h^{-1}(x_0)\Sigma'(x_0)$ into (\ref{localcoord}),
we get
\begin{eqnarray}
x&=&H(1+\xi^aT_{d+1,a})\Sigma(x_0)\nonumber\\
 &=&H(1+\xi^ah(x_0)T_{d+1,a}h^{-1}(x_0))\Sigma'(x_0)\nonumber\\
 &=&H(1+\rho^{({\rm vec})}_{ab}(h(x_0))\xi^bT_{d+1,a})\Sigma'(x_0),
\label{xitrans}
\end{eqnarray}
where $\rho^{({\rm vec})}_{ab}$ is a representation matrix
of vector representation.
We can read in eq.(\ref{xitrans}) that
a local coordinate $\xi'^a$ defined with the section $\Sigma'$ is
$\xi'^a=\rho^{({\rm vec})}_{ab}(h(x_0))\xi^b$
and, certainly, the coordinate is rotated by
the local rotation group $H=SO(d)$ as a vector.

A complete system $\phi^{(r)}_b(x)$ of
a field on $\cal M$ with spin $s$
is given by
\begin{equation}
\phi^{(r)}_b(x)=\rho^{(r)}_{bA}(\Sigma(x))\zeta^A,
\label{mode}
\end{equation}
where $b$ is a spin index
and $r$, which expresses a representation
under the isometry $G$, runs over all representations of $G$
which contain the representation $s$
when they are decomposed into representations of $H$.
$\rho^{(r)}_{bA}$ is a restriction of a representation matrix $\rho^{(r)}_{BA}$
whose former index belongs to $s$.
$\zeta^A$ is constant vector in representation $r$.
Eq.(\ref{mode}) claims that harmonics on $S^d$ are obtained by
projecting constant fields $\zeta^A$ on $R^{d+1}$ into $S^d$.

The fact that the field $\phi^{(r)}_b$ has spin $s$ is confirmed
by observing a transformation law of $\phi^{(r)}_b(x)$ under
a replacement of the section,
which is equivalent to a replacement of the vielbein.
If we replace a section $\Sigma(x)$ by $\Sigma'(x)=h(x)\Sigma(x)$,
the field $\phi^{(r)}_b$ get a $H$ rotation
as a representation $s$ as follows:
\begin{equation}
\phi'^{(r)}_b(x)=\rho^{(r)}_{bA}(\Sigma'(x))\zeta^A
              =\rho^{(s)}_{bc}(h(x))\rho^r_{cA}(\Sigma(x))\zeta^A
              =\rho^{(s)}_{bc}(h(x))\phi^{(r)}_c(x).
\end{equation}

Under an isometry $g\in G$, a point $x\in{\cal M}$ is moved
by $x\rightarrow x'=xg$
and the section is transformed as
\begin{equation}
\Sigma(x')=\Sigma(xg)=h_g(x)\Sigma(x)g,
\end{equation}
where $h_g(x)$ is an element of $H$ depending upon $g$ and $x$.
If we want to remain the vielbein unchanged,
further local rotation $h_g^{-1}(x)$
should be carried out.
Taking account of it,
we obtain the following transformation law of a field $\phi^{(r)}_b(x)$
under an isometry rotation $g$:
\begin{eqnarray}
\phi^{(r)}_b(x)\rightarrow\phi'^{(r)}_b(x)
&=&\rho^{(s)}_{bc}(h^{-1}_g(x))\phi(xg)\nonumber\\
&=&\rho^{(s)}_{bc}(h^{-1}_g(x))\rho^{(r)}_{cA}(\Sigma(xg))\zeta^A\nonumber\\
&=&\rho^{(s)}_{bc}(h^{-1}_g(x))\rho^{(r)}_{cA}
                       (h_g(x)\Sigma(x)g)\zeta^A\nonumber\\
&=&\rho^{(r)}_{cA}(\Sigma(x))\rho^{(r)}_{AB}(g)\zeta^B.
\end{eqnarray}
This equation shows that the isometry rotation
is achieved by a $G$ action on the constant vector $\zeta^A$.

%%%%%%%%%%%%%%%%%%%%%%%%%%%%%%%%%%%%%%%%%%%%%%%%%%%%%%%%%%%%%%%%%%%%%
Finally, we give an expression of a covariant derivative.
Let $x_1$ denote a point on $\cal M$ which is specified
by a coordinate $\xi^a$ on a tangent space at a point $x_0\in{\cal M}$.
The relation among $x_0$, $x_1$ and $\xi^a$ is
\begin{equation}
x_1=H(1+\xi^aT_{d+1,a})\Sigma(x_0).
\end{equation}
Acting a mapping $\Sigma$ on this equation, we get
\begin{equation}
\Sigma(x_1)=\Sigma((1+\xi^aT_{d+1,a})\Sigma(x_0))
           =(1+a^{ab}T_{ab})(1+\xi^aT_{d+1,a})\Sigma(x_0),
\end{equation}
where $a^{ab}$ is a matrix depending on $\xi^a$ linearly.
By using $\Sigma(x_1)-\Sigma(x_0)=\xi^a\partial_a\Sigma(x)$,
we get
\begin{equation}
(\xi^a\partial_a-a^{ab}T_{ab})\Sigma(x)=\xi^aT_{d+1,a}\Sigma(x),
\end{equation}
and this equation suggests that
$\omega_a{}^{bc}\equiv\partial a^{bc}/\partial\xi^a$ and
$T_{d+1,a}$ correspond to a spin connection
and a covariant derivative on $\cal M$ respectively.
The normalization (\ref{Tnormal}) means that the radius of the $S^d$ is unity.
If the radius is $R$, the covariant derivative should be rescaled as
\begin{equation}
D_a=\frac{1}{R}T_{d+1,a}.
\end{equation}

%%%%%%%%%%%%%%%%%%%%%%%%%%%%%%%%%%%%%%%%%%%%%%%%%%%%%%%%%%%%%%%%%%%%%
\section{Global Supersymmetries on $AdS_5\times S^5$}\label{global}

The parameters $\eta^\pm$ giving global supersymmetry transformations
are defined as solutions of differential equations $\delta\psi_M^\pm=0$.
So we should solve the following differential equation
on $AdS_5\times S^5$ background.
\begin{equation}
D_M\eta^\pm=\pm\frac{ia}{5!}
             F_{M_1\ldots M_5}
             \Gamma^{M_1\ldots M_5}
             \Gamma_M\eta^\pm.
\label{difeq}
\end{equation}
In our convention, constant $a$ is
\begin{equation}
a=\frac{\pi^3g_{\rm str}l_s^4}{2}.
\label{constais}
\end{equation}
The five form field strength $F_{M_1\cdots M_5}$ is self-dual and
$F_{12345}=F_{\ol1\ol2\ol3\ol4\ol5}$ in the local Lorentz frame.
Let us refer this common value by $F_5$.
In our convention, the quantization of $p+2$ form gauge flux $F_{p+2}$
is  expressed as
\begin{equation}
\oint F_{p+2}\in 2\pi{\bf Z}.
\end{equation}
Therefore, using the common radius $r_0$ of $AdS_5$ and $S^5$,
$F_5$ is given by
\begin{equation}
F_5=\frac{2\pi N}{c_5r_0^5},
\end{equation}
where $c_5$ is a volume of five dimensional unit sphere.

For the following argument,
it is convenient to decompose the dimension $10$ to two $5$
corresponding to $AdS_5$ and $S^5$ respectively.
Under the decomposition,
the ten dimensional gamma matrices are decomposed as follows:
\begin{eqnarray}
\Gamma^{\ol a}&=&\sigma_y\otimes{\bf 1}_4\otimes\gamma_M^{\ol a},
\quad(\ol a=\ol1,\ol2,\ol3,\ol4,\ol5),\\
\Gamma^a&=&\sigma_x\otimes\gamma_E^a\otimes{\bf 1}_4,
\quad(a=1,2,3,4,5),\\
\Gamma^{11}&=&\sigma_z\otimes{\bf 1}_4\otimes{\bf 1}_4.
\end{eqnarray}
where ${\bf1}_4$ is a $4\times4$ unit matrix
and $\gamma_M^{\ol a}$ and $\gamma_E^a$ are
five dimensional Minkowski and Euclidean gamma matrices, respectively.
Furthermore, the left handed Weyl spinors $\eta^\pm$
are decomposed as follows:
\begin{equation}
\eta^\pm=\uparrow\otimes\eta_E^\pm\otimes\eta_M^\pm,
\end{equation}
where $\uparrow$ is two component spinor $(1,0)^T$
and $\eta_E^\pm$ and $\eta_M^\pm$ are four component $SO(5)$
and $SO(1,4)$ spinors respectively.
Furthermore, $\Gamma^{M_1\cdots M_5}$ is decomposed as
\begin{equation}
\Gamma^{\ol1\ol2\ol3\ol4\ol5}=i\sigma_y\otimes{\bf1}_4\otimes{\bf1}_4,\quad
\Gamma^{12345}=\sigma_x\otimes{\bf1}_4\otimes{\bf1}_4,
\end{equation}
and (\ref{difeq}) is rewritten in a form
\begin{equation}
D^{\ol a}\eta^\pm=\pm iaF_5\left[(\sigma_z+1)\otimes{\bf1}_4
                     \otimes\gamma^{\ol a}_M\right]\eta^\pm,\quad
D^a\eta^\pm=\pm iaF_5\left[(\sigma_z+1)\otimes\gamma^a_E
                     \otimes{\bf1}_4\right]\eta^\pm.
\end{equation}
These equations hold if $\eta_E^\pm$ and $\eta_M^\pm$
depend only upon $x^a$ and $x^{\ol a}$ respectively,
and they are solutions of the following differential equations:
\begin{equation}
D_{\ol a}\eta_M^\pm=\pm2iaF_5\gamma^{\ol a}_M\eta_M^\pm,\quad
D_a\eta_E^\pm=\pm2iaF_5\gamma^a_E\eta_E^\pm.\label{Edifeq}
\end{equation}

First, let us focus on $\eta_E^\pm$.
We should solve the second equation in (\ref{Edifeq}).
In five dimensional space $S^5$, $\eta_E^\pm$ belongs
to $\bf4$ of rocal rotation group $SO(5)$.
According to the argument in the last section,
solving the differential equation is
equivalent to finding $SO(6)$ representation $r$
satisfying
\begin{equation}
(T^{(r)}_{6i})_{aA}\rho^{(r)}_{AB}(\zeta^\pm_E)_B
     =\pm\frac{i}{2}(\gamma_E^i)_{ab}\rho^{(r)}_{bB}(\zeta^\pm_E)_B,
\label{repcond}
\end{equation}
where we use the explicit value (\ref{constais}).
If the representation is found, the spinor fields $\eta^\pm_E$ on $S^5$
is expressed with representation matrix $\rho^{(r)}_{AB}$ as follows:
\begin{equation}
(\eta^\pm_E)_a=\rho^{(r)}_{aA}(\Sigma(x^a))(\zeta^\pm_E)_A.
\end{equation}
It is easy to show that the equation (\ref{repcond}) holds
if $r={\bf4}$ for $\eta^+_E$ and $r=\ol{\bf4}$ for $\eta^-_E$.
Indeed, if $\zeta^\pm$ are Dirac spinors,
which belong to ${\bf4}\oplus\ol{\bf4}$,
the $SO(6)$ generator in (\ref{repcond}) is
\begin{equation}
T_{6i}^{({\bf4}\oplus\ol{\bf4})}
=\frac{1}{2}\Gamma_E^{6i}=\frac{i}{2}(\sigma_z\otimes\gamma_E^i),
\label{t6i}
\end{equation}
where we define six dimensional gamma matrices as follows:
\begin{equation}
\Gamma_E^i=\sigma_x\otimes\gamma_E^i\quad(i=1,\ldots,5),\quad
\Gamma_E^6=\sigma_y\otimes{\bf1}_4,\quad
\Gamma_E^7=\sigma_z\otimes{\bf1}_4.
\end{equation}
Inserting (\ref{t6i}) into (\ref{repcond}), we obtain the condition
for $\zeta_E^\pm$:
\begin{equation}
(\sigma_z\otimes{\bf1}_4)\zeta_E^\pm=\pm\zeta_E^\pm.
\label{psiEcond}
\end{equation}
This condition implies that
$r={\bf 4}$ for $\eta_E^+$ and $r={\ol{\bf 4}}$ for $\eta_E^-$.

Identically, 
we can obtain solutions for $\eta_M^\pm$.
We define six dimensional gamma matrices with signature
$(-,+,+,+,+,-)$ as follows:
\begin{equation}
\Gamma_M^{\ol a}=\sigma_x\otimes\gamma_M^{\ol a}
\quad({\ol a}=\ol1,\ldots,\ol5),\quad
\Gamma_M^{\ol6}=i\sigma_y\otimes{\bf1}_4,\quad
\Gamma_M^{\ol7}=\sigma_z\otimes{\bf1}_4.
\end{equation}
Then, $\eta_M^\pm$ is expressed as
\begin{equation}
(\eta_M^\pm)_a=\rho^{({\bf4}\oplus\ol{\bf4})}_{aA}
(\Sigma(x^{\ol a}))(\zeta_M^\pm)_A,
\end{equation}
with six dimensional Weyl spinor $\zeta_M^\pm$ satisfying
\begin{equation}
(\sigma_z\otimes{\bf1}_4)\zeta^\pm_M=\pm\zeta^\pm_M.\label{psiMcond}
\end{equation}
Combining (\ref{psiEcond}) and (\ref{psiMcond}),
we obtain Weyl spinors with complex $16$ independent components:
\begin{equation}
\eta^+=\uparrow\otimes\rho^{(\bf4)}(\Sigma(x^a))\zeta_E^+
                 \otimes\rho^{(\bf4)}(\Sigma(x^{\ol a}))\zeta_M^+,\quad
\eta^-=\uparrow\otimes\rho^{(\ol{\bf4})}(\Sigma(x^a))\zeta_E^-
                 \otimes\rho^{(\ol{\bf4})}(\Sigma(x^{\ol a}))\zeta_M^-.
\label{etais}
\end{equation}
Symmetries connected with these parameters $\eta^\pm$
consists of rigid supersymmetries $Q$
and superconformal symmetries $S$.
To pick up rigid supersymmetry transformations $Q$,
we should impose the invariance under the parallel transports $P_{\ol a}$
on $\eta^\pm$.
As we have already mentioned,
$P_{\ol a}$ corresponds to the generators $T_{+{\ol a}}$ of $SO(2,4)$
and the isometry rotation is realized by a rotation of $\zeta^\pm_M$.
Therefore, $\zeta^\pm_M$ for rigid supersymmetries $Q$ should satisfy
\begin{equation}
T_{+{\ol a}}\zeta^\pm_M=0.
\end{equation}
Because the generator is represented as
\begin{equation}
T_{+{\ol a}}=\frac{1}{2}(\Gamma_M^{\ol6}-\Gamma_M^{\ol5})\Gamma_M^{\ol a}
=\frac{1}{2}\Gamma_M^{\ol6\ol a}({\bf1}_8-\Gamma_M^{\ol5\ol6}),
\end{equation}
the invariance under $T_{+{\ol a}}$ means that
the spinors have a positive charge
under $x^{\ol5}$-$x^{\ol6}$ rotation.
\begin{equation}
\Gamma_M^{\ol5\ol6}\zeta^\pm_M=-(\sigma_z\otimes\gamma_M^{\ol5})\zeta^\pm_M
=+\zeta^\pm_M.\label{-15charge}
\end{equation}

The charge conjugation operation in ten dimension is expressed as
\begin{equation}
(\uparrow\otimes\eta_E\otimes\eta_M)_c
=({\bf1}_2\otimes C_5^\dagger\otimes\gamma^{\ol1}C_5^\dagger)
 (\uparrow\otimes\eta_E\otimes\eta_M),
\end{equation}
where $C_5$ is five dimensional charge conjugation matrix
which satisfies the following equations:
\begin{equation}
C_5\gamma_a=+\gamma_a^TC_5,\quad
C_5^T=-C_5.
\end{equation}
Therefore, we can set
\begin{equation}
\eta^-_E=+i(\eta^+_E)_c,\quad
\eta^+_E=-i(\eta^-_E)_c,\quad
\eta^-_M=-i(\eta^+_M)_c,\quad
\eta^+_M=+i(\eta^-_M)_c,
\label{5dimcg}
\end{equation}
where a charge conjugation $\psi_c$ of a five dimensional spinor $\psi$ is
defined by
\begin{equation}
\psi_c=C_5^\dagger\psi^* \quad\mbox{(Euclidean)},\quad
\psi_c=\gamma^{\ol1}C_5^\dagger\psi^* \quad\mbox{(Minkowski)}.
\end{equation}
The relations (\ref{5dimcg}) are equivalent
to the statement that two six dimensional spinors
\begin{equation}
\eta'_E=\left(\begin{array}{c} \eta_E^+ \\ \eta_E^- \end{array}\right),\quad
\eta'_M=\left(\begin{array}{c} \eta_M^+ \\ \eta_M^- \end{array}\right),
\label{6spinor}
\end{equation}
are Majorana spinors.
%%%%%%%%%%%%%%%%%%%%%%%%%%%%%%%%%%%%%%%%%%%%%%%%%%%%%%%%%%%%%%%%%%%%%
\section{Quark Configuration}
In terms of a brane configuration,
an external quark in the conformal field theory
is represented by a semi-infinite open string
attached on the three brane.
Its world volume is an intersection
of three dimensional flat plane containing $x^{\ol1}$, $x^{\ol5}$
and $x^{\ol6}$ axes and $AdS_5$ (\ref{pseudosphere}).
Let ${\cal M}_w\subset AdS_5$ denote the world volume.
To be precise, the world volume
is not just ${\cal M}_w$ but 
${\cal M}_w\otimes P$ where $P$ is a point on $S^5$.
We suppose the point $P$ to be $He\in H\backslash G$.
The manifold ${\cal M}_w$ is a copy of $AdS_2$ and is expressed as a coset space
${\cal M}_w=H\backslash(HG_w)$ where $G_w$ is $SO(2,1)\subset G=SO(2,4)$
generated by $T_{\ol6\ol1}$,
$T_{\ol6\ol5}$ and their commutator $T_{\ol5\ol1}$.
A point $x\in{\cal M}_w$ is identified with a set $Hg_w$
where $g_w$ is a element of $G_w$.
To introduce a local frame on $G_w$,
let us take a section $\Sigma$ satisfying a condition
\begin{equation}
\Sigma(Hg_w(\tau,\sigma))\in G_w.\label{choice}
\end{equation}
To determine an embedding of the tangent space of ${\cal M}_w$ in $AdS_5$,
taking variation of $x=H\Sigma(x)$ and we get
\begin{equation}
\delta x=H\delta\Sigma(x)
=H(\epsilon_1T_{\ol6\ol1}+\epsilon_2T_{\ol6\ol5}
                         +\epsilon_3T_{\ol1\ol5})\Sigma(x)
        =H(\epsilon_1T_{\ol6\ol1}+\epsilon_2T_{\ol6\ol5})\Sigma(x),
\label{variation}
\end{equation}
where $\epsilon_1$, $\epsilon_2$ and $\epsilon_3$
are infinitesimal coefficients.
Eq.(\ref{variation}) implies that in our choice of the section (\ref{choice}),
tangent space of ${\cal M}_w$ is spanned by $\xi^{\ol1}$ and $\xi^{\ol5}$.
Therefore, using (\ref{flatbps}),
we obtain the conditions for unbroken supersymmetries.
\begin{equation}
({\bf1}_4\otimes\gamma_M^{\ol1\ol5})(\eta_E^+\otimes\eta_M^+)
=(\eta_E^-\otimes\eta_M^-),\quad
({\bf1}_4\otimes\gamma_M^{\ol1\ol5})(\eta_E^-\otimes\eta_M^-)
=(\eta_E^+\otimes\eta_M^+).
\label{stringcond}
\end{equation}
In terms of the six dimensional Majorana spinors (\ref{6spinor}),
these are expressed as
\begin{equation}
\Gamma_E^6\eta'_E=\eta'_E,\quad
i\Gamma_M^{\ol6\ol1\ol5}\eta'_M=\eta'_M.
\label{stringunbroken}
\end{equation}
These equations are interpreted as follows.
The world sheet of the string is direct product of ${\cal M}_w\in AdS_5$ and
the point $P$ on $S^5$.
${\cal M}_w$ is an intersection of the $AdS_5$ and a three dimensional plane
lying along $x^{\ol1}$, $x^{\ol5}$ and $x^{\ol6}$
and $P$ is an intersection of the $S^5$ and a line along $x^6$.
The equations (\ref{stringunbroken}) are regarded
as supersymmetry breaking pattern due to the three dimensional plane
and the line.

In section \ref{sec:baryon}, we will construct baryon configuration.
For the purpose, it is convenient to introduce the following
parameterization on ${\cal M}_w$.
\begin{equation}
\Sigma(g)=\exp(tT_{\ol1+})\exp(rT_{\ol5\ol6}),\quad
g\in G_w.
\label{ads2coord}
\end{equation}
Eq.(\ref{ads2coord}) fixes not only coordinate on ${\cal M}_w$
but also a local frame on ${\cal M}_w$.
The relation between the Cartesian coordinates
$x^{\ol1}$, $x^{\ol5}$ and $x^{\ol6}$
and worldsheet coordinates $t$ and $r$ is
\begin{equation}
\left(\begin{array}{c} x^{\ol1} \\ x^{\ol5} \\ x^{\ol6} \end{array}\right)
=
\left(\begin{array}{ccc} 1 & t & t \\
                         t & 1+t^2 & t^2 \\
                         -t & -t^2 & 1-t^2 \end{array}\right)
\left(\begin{array}{ccc} 1 \\
                         & \cosh r & \sinh r \\
                         & \sinh r & \cosh r \end{array}\right)
\left(\begin{array}{c} 0 \\ 0 \\ r_0 \end{array}\right).
\end{equation}
Then, metric on ${\cal M}_w=AdS_2$ is
\begin{equation}
ds^2=r_0^2(-e^{2r}dt^2+dr^2).
\end{equation}
In this metric, energy of quark configuration is
\begin{equation}
E_{\rm str}=T_{\rm FS}r_0^2\int e^rdr.
\label{estr}
\end{equation}
For simplicity, let us restrict our argument on a time slice $t=0$.
Then, using a section (\ref{ads2coord}), $\eta^\pm$ is given by
\begin{equation}
\eta^\pm(r)=\uparrow\otimes\zeta^\pm_E\otimes
e^{\mp r\gamma^{\ol5}/2}\zeta^\pm_M.
\label{stringzeta}
\end{equation}
From (\ref{stringcond}) and (\ref{stringzeta}),
we obtain the following constraints
on $\zeta^\pm_E$ and $\zeta^\pm_M$:
\begin{equation}
({\bf1}_4\otimes\gamma_M^{\ol1\ol5})(\zeta_E^+\otimes\zeta_M^+)
=(\zeta_E^-\otimes\zeta_M^-),\quad
({\bf1}_4\otimes\gamma_M^{\ol1\ol5})(\zeta_E^-\otimes\zeta_M^-)
=(\zeta_E^+\otimes\zeta_M^+).
\label{zetacond}
\end{equation}

%%%%%%%%%%%%%%%%%%%%%%%%%%%%%%%%%%%%%%%%%%%%%%%%%%%%%%%%%%%%%%%%%%%%%
\section{Local BPS Condition}
In this section, we explain a condition for the brane orientation
not to break too many supersymmetries in a flat background.
It has been already discussed in \cite{BIstring1,BIstring2} in detail.

An electric field on a D-brane can be regarded as fundamental strings bound
on the D-brane
because $F_{\mu\nu}$ couples to NS-NS two form field $B_{\mu\nu}$\cite{mixed}.
A fundamental string current carried by D5-brane is
equal to the dual field strength $\wt F_{\mu\nu\rho\sigma}$
defined by
\begin{equation}
\frac{1}{2\pi}\wt F_{\mu\nu\rho\sigma}
=\frac{1}{2}\epsilon_{\mu\nu\rho\sigma\alpha\beta}
\frac{\partial S}{\partial F_{\alpha\beta}},
\end{equation}
because the field strength $F_{\mu\nu}$ always appears in the D-brane action $S$
as a combination $F_{\mu\nu}-B_{\mu\nu}$.
From this viewpoint, dual field $\wt F_{\mu\nu\rho\sigma}$ is
identified with the density of the strings.
Namely, the quantized electric flux
\begin{equation}
\frac{1}{2\pi}\oint_C\wt F_4\in{\bf Z},
\end{equation}
is equal to the number of strings go through the closed surface $C$.

Using this dual gauge field $\wt F_{\mu\nu\rho\sigma}$,
the constraint for unbroken supersymmetries on a D5-brane is
\begin{equation}
\frac{1}{T'_{D5}}S_{\alpha\beta\gamma\delta\epsilon\zeta}
\left(\frac{T_{FS}}{4!2!}
\frac{1}{2\pi}\wt F^{\alpha\beta\gamma\delta}\Gamma^{\epsilon\zeta}
\mp i\frac{T_{D5}}{6!}\Gamma^{\alpha\beta\gamma\delta\epsilon\zeta}\right)
\eta^\pm=\eta^\mp,
\label{localBPS}
\end{equation}
where $S_{\alpha\beta\gamma\delta\epsilon\zeta}$ is
anti-symmetric tensor expressing the orientation of the D5-brane
normalized by $S_{\alpha\beta\gamma\delta\epsilon\zeta}^2/6!=1$,
and $T_{D5}'$ is modified tension of the D5-brane,
which is an energy per unit volume of the D5-brane
containing one of the electric field on the brane.
We can reach to the expression (\ref{localBPS}) by the following way.
Let us consider a situation in which
the D5-brane and the fundamental strings exist separately.
For concreteness, we suppose that
the D5-brane and the strings spread
along $dx^0\wedge\cdots\wedge dx^5$ and $dx^0\wedge dx^1$ respectively.
% orientation of electric field is along $x^1$.
If these branes make a bound state, the resultant configuration is
a D5-brane with an electric field on it.
The unique non zero element of the dual gauge field strength
$\wt F_{\mu\nu\rho\sigma}$ is $\wt F_{2345}$.
Let us compactify five directions $x^1$, $x^2$, $x^3$, $x^4$ and $x^5$ on $T^5$.
Then, energies of the D5-brane $M_{D5}$
and the string $M_{FS}$ get finite:
\begin{equation}
M_{FS}=T_{FS}V_5\frac{\wt F_4}{2\pi},\quad
M_{D5}=T_{D5}V_5,
\end{equation}
where $V_5$ is a volume of the torus $T^5$
and $\wt F_4=\sqrt{\wt F_{\mu\nu\rho\sigma}^2/4!}$.
Carrying out a series of dualities $T_4$, $T_5$, $S$ and $T_3$ successively
($S$ and $T_i$ mean S-duality and T-duality along $x^i$ respectively),
the D5-brane and the fundamental strings transformed into
D2-brane wrapped along $x^1$ and $x^2$ and ones wrapped along $x^1$ and $x^3$.
Although the compactification radii are changed
by means of the duality transformations,
the masses of the branes are invariant.
The D2-branes combine into bound state, which is a D2-brane wrapped
along the slanted direction:
\begin{equation}
dx^0\wedge dx^1\wedge\left(\frac{M_{D5}}{M_{\rm tot}}dx^2
                          +\frac{M_{FS}}{M_{\rm tot}}dx^3\right).
\end{equation}
where $M_{\rm tot}$ is
the mass of the bound state obtained by Pythagoras' theorem:
\begin{equation}
M_{\rm tot}=\sqrt{M_{FS}^2+M_{D5}^2}.
\label{Mprime}
\end{equation}
Therefore, the constraint for unbroken supersymmetries on the D2-brane is
\begin{equation}
\left(\frac{M_{D5}}{M_{\rm tot}}\Gamma^{012}
     +\frac{M_{FS}}{M_{\rm tot}}\Gamma^{013}\right)\eta_L=\eta_R,
\end{equation}
where $\eta_L$ and $\eta_R$ are
supersymmetry parameters of Type-IIA theory.
If we trace the series of dualities in the opposite order,
we obtain the constraint (\ref{localBPS}).
Furthermore, from (\ref{Mprime}),
we get the modified tension of the D5-brane\cite{mixed}:
\begin{equation}
T_{D5}'=T_{D5}\sqrt{1+f^2},\quad
f=\frac{T_{FS}}{T_{D5}}\frac{\wt F_4}{2\pi}.
\end{equation}

Next, let us consider a D5-brane lying along $x^1,\ldots,x^5$
on which a semi-infinite open string along $x^6$ is attached.
The unbroken supersymmetries in this configuration
is determined by constraints
\begin{equation}
\Gamma^{06}\eta^\pm=\eta^\mp,\quad
\Gamma^{012345}\eta^\pm=\pm i\eta^\mp.
\label{fsd5}
\end{equation}
The first equation comes from the fundamental string and another one comes from
the D5-brane.
If the string tension is not negligible,
the D5-brane is deformed by it
and, simultaneously, we cannot neglect the electric field
on the D5-brane.
In this case, we should use the Born-Infeld action
to find the brane configuration.
However, it is known that if the configuration does not break
supersymmetries determined by (\ref{fsd5}),
it automatically satisfies the equation of motion
obtained from the Born-Infeld action.
Therefore, we can find the solution by means of BPS the condition.
If we use the worldvolume coordinate $\sigma^i=x^i$, ($i=0,\ldots,5$),
the brane configuration is specified by giving a function
$x^6(\sigma)$ and an electric field $E_i(\sigma)$.
Due to a rotational symmetry, we can
assume that the direction of $E_i$ and $\nabla_ix^6$ is
along $x^5$.
It is convenient to define an angle $\phi$ by
$\partial_5x^6=\tan\phi$.
Then, constraint for unbroken supersymmetries is given by
\begin{equation}
\frac{1}{\sqrt{1+f^2}}
(fn_\mu\Gamma^{0\mu}\eta^\pm
\mp in_\mu\Gamma^{01234\mu}\eta^\pm)=\eta^\mp.
\end{equation}
where $n_\mu$ is unit vector with $n_5=\cos\phi$ and $n_6=\sin\phi$.
Using (\ref{fsd5}), we obtain
\begin{equation}
\frac{1+f\Gamma^{56}}{\sqrt{1+f^2}}
\eta^\pm=\exp(\phi\Gamma^{56})\eta^\pm.
\label{56eta}
\end{equation}
For the brane configuration to be a BPS state,
the constraint (\ref{56eta}) must not give any new constraint.
Because eigenvalues of $\Gamma^{56}$ are $\pm i$,
$f$ and $\theta$ should satisfy the following relation.
\begin{equation}
\arg(1+if)=\phi.
\label{arg1pif}
\end{equation}
This condition is equivalent to
\begin{equation}
f=\partial_5X^6,
\label{fgradient}
\end{equation}
which has been already given in \cite{BIstring1,BIstring2}.
The flux density $f$ is easily obtained
by means of flux conservation law,
and we can easily obtain the brane configuration
by integrating (\ref{fgradient}).

%%%%%%%%%%%%%%%%%%%%%%%%%%%%%%%%%%%%%%%%%%%%%%%%%%%%%%%%%%%%%%%%%%%%%
%%%%%%%%%%%%%%%%%%%%%%%%%%%%%%%%%%%%%%%%%%%%%%%%%%%%%%%%%%%%%%%%%%%%%
\section{Baryon Configuration}\label{sec:baryon}
At last, we are ready to calculate the binding energy of a baryon.
In terms of brane configuration,
a baryon is expressed as a D5-brane wrapped around $S^5$
on which $N$ strings with the same orientation are attached.
This configuration is embedded in ${\cal M}_w\times S^5$.
We define a coordinate $\theta^1,\ldots\theta^5$ on $S^5$ such that
$\theta^5$ represents an angle from the opposite point to the string endpoint
and the other four parameters $\theta^1$, $\theta^2$, $\theta^3$ and $\theta^4$
are angular variables parameterizing $S^4\subset S^5$ with fixed $\theta^5$.
Then, the metric on $AdS_2\times S^5$ is
\begin{equation}
ds^2=r_0^2(-e^{2r}dt^2+dr^2)
+r_0^2((d\theta^5)^2+\sin^2\theta s_{ij}d\theta^id\theta^j),
\end{equation}
where $s_{ij}$ is a metric on a four dimensional unit sphere.
Due to a symmetry, brane configuration is expressed by unique function:
\begin{equation}
r=r(\theta^5).\label{Bconfig}
\end{equation}
If we use $(t,\theta^1,\ldots,\theta^5)$ as world volume coordinates,
an induced metric on the D5-brane is
\begin{equation}
h_{ij}=r_0^2\left(\begin{array}{ccc}
           -e^{2r} \\
           & \sin^2\theta^5 s_{ij} \\
           && 1+\left(\frac{dr}{d\theta^5}\right)^2 \end{array}\right).
\end{equation}
Therefore, the energy of the D5-brane is
\begin{equation}
E_{D5}
=\int d^5\theta \sqrt{-\det h_{ij}}T_{D5}'
=T_{D5}r_0^6\int d\theta^5e^r\sqrt{1+\left(\frac{dr}{d\theta^5}\right)^2}
  c_4\sin^4\theta^5\sqrt{1+f^2},
\label{ed5}
\end{equation}
where $c_4=8\pi^2/3$ is a volume of the four dimensional unit sphere.
We should find the function $r(\theta^5)$
giving the minimum value of the energy (\ref{ed5}).
The Euler-Lagrange equation for function $r(\theta^5)$ is
obtained from (\ref{ed5}) as follows:
\begin{equation}
\frac{d}{d\theta^5}\left(\sin\phi\sin^4\theta^5\sqrt{1+f^2}\right)
=\cos\phi\sin^4\theta^5\sqrt{1+f^2}
\label{eleq}
\end{equation}
where we have introduced an angular variable $\phi$ defined by
\begin{equation}
\tan\phi=\frac{dr}{d\theta^5}.
\end{equation}
To define the normalized flux $f$
in the expression (\ref{eleq}),
we must not use the canonical momentum
\begin{equation}
\frac{\wt F^{\rm(can)}_{\mu\nu\rho\sigma}}{2\pi}
=\frac{1}{2}\epsilon_{\mu\nu\rho\sigma\alpha\beta}
\frac{\partial S}{\partial F_{\alpha\beta}},\quad
S=S_{\rm BI}+S_{\rm CS},
\end{equation}
of electric field,
because it is not invariant
under the gauge transformation of R-R four form field $\delta C_4=d\Lambda_3$.
Instead, we have to use the gauge invariant one:
\begin{equation}
\frac{\wt F^{\rm(GI)}_{\mu\nu\rho\sigma}}{2\pi}
=\frac{1}{2}\epsilon_{\mu\nu\rho\sigma\alpha\beta}
\frac{\partial S_{BI}}{\partial F_{\alpha\beta}}
=\frac{\wt F^{\rm(can)}_{\mu\nu\rho\sigma}}{2\pi}
+\frac{C_{\mu\nu\rho\sigma}}{2\pi}.
\label{fgiis}
\end{equation}
Due to an equation of motion,
canonical flux $\wt F^{\rm(can)}_{\mu\nu\rho\sigma}$ is conserved.
Therefore, by integrating (\ref{fgiis}), we obtain
\begin{equation}
\frac{1}{2\pi}\int_{S^4}\wt F^{\rm GI}_4=\frac{1}{2\pi}\int_{S^4}C
=\frac{1}{2\pi}\int_{D_5}F_5,
\label{intfgi}
\end{equation}
where $S^4$ is a section of $S^5$ with fixed $\theta^5$
and $D^5$ is five dimensional disc with boundary $S^4$.
From the two choice of $D^5$ for given $S^4$,
we take one containing a point $\theta^5=0$.
From eq.(\ref{intfgi}), we can easily obtain $\wt F^{\rm(GI)}_4$ as follows:
\begin{equation}
\frac{1}{2\pi}\wt F^{\rm(GI)}_4(\theta^5)
=\frac{N}{r_0^4\pi^3}\frac{1}{\sin^4\theta^5}
\int_0^{\theta^5}\sin^4\theta^5d\theta^5,
\end{equation}
and $f$ is given by
\begin{equation}
f(\theta^5)=\frac{T_{FS}}{T_{D5}}\frac{\wt F^{\rm(GI)}_4}{2\pi}
=\frac{4}{\sin^4\theta^5}\int_0^{\theta^5}\sin^4\theta^5d\theta^5.
\label{flux}
\end{equation}
To solve the differential equation (\ref{eleq}),
we should find the BPS condition
for the D5-brane orientation on $AdS_5\times S^5$.
In the quark configuration, unbroken supersymmetry satisfies
the conditions (\ref{psiEcond}), (\ref{psiMcond}), (\ref{-15charge})
and (\ref{zetacond}).
If we assume the further breaking of the supersymmetries does not occur
in a creation process of a baryon,
we should find a condition for the D5-brane orientation
not to break the supersymmetries.
On a time slice $t=0$,
let us parameterize $r$-$\theta^5$ plane by
\begin{equation}
\Sigma(r)=\exp(rT_{\ol5\ol6})\in SO(4,2),\quad
\Sigma(\theta^5)=\exp(\theta^5T_{56})\in SO(6).
\label{sectionads5s5}
\end{equation}
(The first equation in (\ref{sectionads5s5})
is the same with (\ref{ads2coord}).)
Then, the parameter fields of unbroken supersymmetries are given by
\begin{equation}
\eta^+(r,\theta^5)=\uparrow\otimes
e^{i\theta^5\gamma^5/2}\zeta^+_E
\otimes e^{-r\gamma^{\ol5}/2}\zeta^+_M,\quad
\eta^-(r,\theta^5)=\uparrow\otimes
e^{-i\theta^5\gamma^5/2}\zeta^-_E
\otimes e^{r\gamma^{\ol5}/2}\zeta^-_M.
\label{paramonads}
\end{equation}
Inserting (\ref{paramonads}) to (\ref{zetacond}),
we obtain the following relation:
\begin{equation}
\Gamma^{\ol1\ol5}\eta^+(r,\theta^5)=
e^{i\theta^5\gamma^5}\eta^-(r,\theta^5),\quad
\Gamma^{\ol1\ol5}\eta^-(r,\theta^5)=
e^{-i\theta^5\gamma^5}\eta^+(r,\theta^5).
\label{15psi}
\end{equation}
Furthermore, the following equations hold due to eqs.(\ref{psiEcond}),
(\ref{psiMcond}) and (\ref{-15charge}):
\begin{equation}
\Gamma^{\ol55}\eta^\pm=\pm i\gamma_E^5\eta^\pm,\quad
\Gamma^{1234}\eta^\pm=\gamma_E^5\eta^\pm.
\label{551234}
\end{equation}
The tangent space of D5-brane in the baryon configuration (\ref{Bconfig}) is
$dt\wedge
d\theta^1\wedge
d\theta^2\wedge
d\theta^3\wedge
d\theta^4\wedge
(\cos\phi dr+\sin\phi d\theta^5)$.
Therefore, a constraint for unbroken supersymmetries is given by
\begin{equation}
\frac{1}{\sqrt{1+f^2}}n_\mu(\mp i\Gamma^{\ol11234\mu}+f\Gamma^{\ol1\mu})
\eta^\pm=\eta^\mp,
\end{equation}
where $n_\mu$ is a unit vector with $n_{\ol5}=\cos\phi$ and $n_5=\sin\phi$.
Using (\ref{15psi}) and (\ref{551234}), this is rewritten as
\begin{equation}
\exp\left[\mp i(\arg(1+if)+\phi-\theta^5)\gamma_E^5\right]
\eta^\pm=\eta^\pm.
\label{baryonBPScond}
\end{equation}
For the D5-brane configuration to be a BPS configuration,
the (\ref{baryonBPScond}) must not give any new constraints on $\eta^\pm$.
Therefore,
the following condition should be satisfied:
\begin{equation}
\arg(1+if)+\phi-\theta^5=0.
\label{bpscond}
\end{equation}
From this condition, we get
\begin{equation}
\frac{dr}{d\theta^5}=\tan\phi
=\frac{\sin\theta^5-f\cos\theta^5}{\cos\theta^5+f\sin\theta^5},\quad
\sin\phi=\frac{\sin\theta^5-f\cos\theta^5}{\sqrt{1+f^2}},\quad
\cos\phi=\frac{\cos\theta^5+f\sin\theta^5}{\sqrt{1+f^2}}.
\label{sincostan}
\end{equation}
Using these equation, the Euler-Lagrange equation (\ref{eleq})
is reduced to the simple equation:
\begin{equation}
\frac{d}{d\theta^5}(f\sin^4\theta^5)=4\sin^4\theta^5,
\end{equation}
and this holds thanks to (\ref{flux}).

\begin{figure}[hbt]
\centerline{\epsfbox{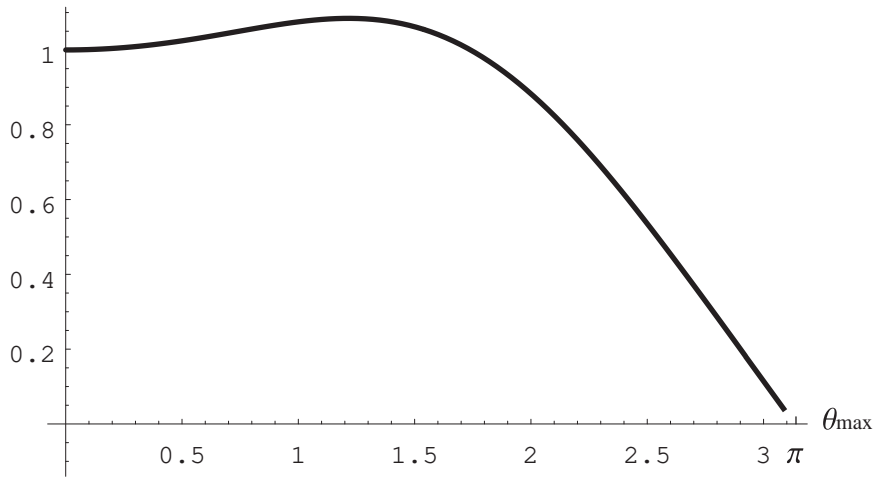}}
\caption{$\theta_{\rm max}$ dependence of
$[NE_{\rm str}(\theta_{\rm max})-E_{D5}(\theta_{\rm max})]
/NE_{\rm str}(\theta_{\rm max}=0)$.}
\label{energyplot.eps}
\end{figure}

Finally, we should calculate the energy.
Near $\theta^5=\pi$, $dr/d\theta^5$ and $f$
diverge and we can neglect the `$1$' in the square roots in
eq.(\ref{ed5}). Therefore, we obtain the following expression
for the energy, which is
identical with $N$ times of (\ref{estr}):
\begin{equation}
E_{D5}=T_{D5}r_0^6\int dr 4\pi^3e^r=NT_{FS}r_0^2\int e^rdr,
\end{equation}
and this confirms that in the asymptotic region with large $r$,
the BI-string is identified with usual fundamental strings.
In the near horizon region with small $r$, we cannot neglect the `1'.
Inserting (\ref{sincostan}) to the expression of the D5-brane energy (\ref{ed5}),
we obtain
\begin{equation}
E_{D5}=NT_{FS}r_0^2\frac{2}{3\pi}
\int e^r\frac{\sin^4\theta^5(1+f^2)}{\cos\theta^5+f\sin\theta^5}d\theta^5.
\label{ed5reduce}
\end{equation}
We can obtain function $r(\theta)$ by integrating
the first equation of (\ref{sincostan}).
And carrying out the integration in (\ref{ed5reduce}),
we obtain the total energy of the baryon configuration.
To pick up the binding energy, we should subtract $N$ times of the energy
of string configuration (\ref{estr}) from (\ref{ed5reduce}).
Because both $E_{\str}$ and $E_{D5}$ diverge, we should introduce
cut off $r_{\rm max}=r(\theta_{\rm max})$.
\begin{equation}
E_{D5}(\theta_{\rm max})=NT_{FS}r_0^2\frac{2}{3\pi}
        \int_0^{\theta_{\rm max}}
   e^r\frac{\sin^4\theta^5(1+f^2)}{\cos\theta^5+f\sin\theta^5}d\theta^5,\quad
E_{FS}(\theta_{\rm max})=T_{FS}r_0^2
        \int_0^{r(\theta_{\rm max})}
         e^r dr.
\end{equation}
The binding energy is obtained by taking a limit:
\begin{equation}
E_{\rm bind}=\lim_{\theta_{\rm max}\rightarrow\pi}
            (NE_{FS}(\theta_{\rm max})-E_{D5}(\theta_{\rm max})).
\end{equation}
The $\theta_{\rm max}$ dependence of 
$NE_{FS}(\theta_{\rm max})-E_{D5}(\theta_{\rm max})$
obtained by numerical integration is plotted in Fig.\ref{energyplot.eps}.
It shows that the binding energy vanishes.
This means that the baryon configuration is a marginal bound state
as is expected from supersymmetry.

%%%%%%%%%%%%%%%%%%%%%%%%%%%%%%%%%%%%%%%%%%%%%%%%%%%%%%%%%%%
\section{Conclusion}

In this paper, we have calculated an energy of a baryon configuration
on $AdS_5\times S^5$.
To solve the equation of motion coming from the Born-Infeld action
(and the Chern-Simons action), we used a BPS condition for a brane orientation,
which guarantees that supersymmetry breaking does not occur
in a baryon creation process.
As a result, it has been shown that
the binding energy vanishes and
it is consistent with what is expected from
supersymmetry.

\section*{Acknowledgment}

This work was supported in part by Grant-in-Aid for Scientific
Research from Ministry of Education, Science and Culture
(\#9110).

%%%%%%%%%%%%%%%%%%%%%%%%%%%%%%%%%%%%%%%%%%%%%%%%%%%%%%%%%%%%%%%%%%%%%
%%%%%%%%%%%%%%%%%%%%%%%%%%%%%%%%%%%%%%%%%%%%%%%%%%%%%%%%%%%%%%%%%%%%%

\end{document}